\documentclass[showpacs,preprintnumbers,amsmath,amssymb,nofootinbib]{revtex4}
\usepackage{graphicx}
\usepackage{dcolumn}
\usepackage{bm,color}

\newcommand{\refeq}[1]{~(\ref{#1})}

\newcommand{\eqd}{\stackrel{d}{=}}
\newcommand{\pslash}{p\!\!\!\!\diagup}
\newcommand{\dslash}{\partial\!\!\!\!\diagup}

\begin{document}

\title{ \textbf{Mass spectrum from stochastic L\évy-Schr\"odinger relativistic equations:\\
possible  qualitative predictions in QCD}}
\author{Nicola \surname{Cufaro Petroni}}
 \email{cufaro@ba.infn.it}
 \affiliation{Department of Mathematics and TIRES, Bari
 University;\\
 INFN Sezione di Bari, \\
via E Orabona 4, 70125 Bari, Italy}
\author{Modesto Pusterla}
 \email{pusterla@pd.infn.it}
 \affiliation{Department of Physics, Padova
 University;\\
 INFN Sezione di Padova, \\
via F.\ Marzolo 8, 35100 Padova, Italy}

\begin{abstract}
\noindent Starting from the relation between the kinetic energy of a free L\évy-Schr\"odinger
particle and the logarithmic characteristic of the underlying stochastic process, we show that it
is possible to get a precise relation between renormalizable field theories and a specific L\évy
process. This subsequently leads to a particular cut-off in the perturbative diagrams and can
produce a phenomenological mass spectrum that allows an interpretation of quarks and leptons
distributed in the three families of the standard model.
%We introduce a modification in the relativistic equations in such a way that (1) the relativistic
%Schr\"odinger equations can always be based on an underlying L\évy process, (2) several families
%of particles with different rest masses can be selected, and finally (3) the corresponding Feynman
%diagrams are convergent when we have at least three different masses.
\end{abstract}


\maketitle

\section{Introduction and notations}\label{intro}

In this note we adopt the space-time relativistic approach of Feynman's propagators (for bosons
and fermions) instead of the canonical Lagrangian-Hamiltonian quantized field theory. The
rationale for this choice is that for the development of our basic ideas the former alternative is
better suited to exhibit the connection between the propagator of quantum mechanics and the
underlying L\évy processes. More precisely, the relativistic Feynman propagators are here linked
to a dynamical theory based on a particular L\évy process: a point, already discussed in a
previous paper~\cite{cufaro09}, which is here analyzed thoroughly with the purpose of deducing its
consequences for the basic interactions among the fundamental constituents, namely quarks,
leptons, gluons, photons etc. To this end we first recall that a L\évy process is a stochastic
process $X(t),\;t\geq0$ on a probability space $(\Omega,\mathcal{F},\mathbb{P})$ such that
\begin{itemize}
    \item $X(0)=0,\quad \mathbb{P}$-q.o.
    \item $X(t)$ has independent and stationary increments: for each $n$ and for very choice of
    $0\leq t_1<t_2<\ldots<t_n<+\infty$ the increments $X(t_{j+1})-X(t_j)$ are independent and $X(t_{j+1})-X(t_j)\eqd
    X(t_{j+1}-X(t_j)$;
    \item $X(t)$ is stochastically continuous: for every $a>0$ and for every $s$
    \begin{equation*}
        \lim_{t\to s}\mathbb{P}\left(|X(t)-X(s)|>a\right)=0.
    \end{equation*}
\end{itemize}
To simplify the notation we will consider in the following only one-dimensional L\évy processes
(an $n$-dimensional extension, however, would not be a very difficult task): it is well
known~\cite{sato,applebaum,cufaro08} that all its laws are infinitely divisible, but we will be
mainly interested in the non stable (and in particular non Gaussian) case. In other words the
characteristic function of the process $\Delta t$-increment is $[\varphi(u)]^{\Delta t/\tau}$
where $\varphi$ is infinitely divisible, but not stable\footnote{A law $\varphi$ is said to be
\emph{infinitely divisible} if for every $n$ it exists a characteristic function $\varphi_n$ such
that $\varphi=\varphi_n^n$; on the other hand it is said to be \emph{stable} when for every $c>0$
it is always possible to find $a>0$ and $b\in\mathbf{R}$ such that
$e^{ibu}\varphi(au)=[\varphi(u)]^c$. Every stable law is also infinitely divisible.}, and $\tau$
is a time scale parameter. The transition probability density $p(2|1)=p(x_2,t_2|x_1,t_1)$ of a
particle moving from the space-time point 1 to 2 then is
\begin{equation}\label{transpdf}
    p(2|1)=\frac{1}{2\pi}\int_{-\infty}^{+\infty}du\,[\varphi(u)]^{(t_2-t_1)/\tau}e^{-iu(x_2-x_1)}
\end{equation}
In analogy with the non relativistic Wiener case, we then obtain for the motion of a free particle
the Feynman propagator $\mathcal{K}(2|1)=\mathcal{K}(x_2,t_2|x_1,t_1)$ as
\begin{equation}\label{propagator}
    \mathcal{K}(2|1)=\frac{1}{2\pi}\int_{-\infty}^{+\infty}du\,[\varphi(u)]^{i(t_2-t_1)/\tau}e^{-iu(x_2-x_1)}
\end{equation}
and the corresponding wave function evolution is
\begin{equation}\label{evolution}
    \psi(x,t)=\int_{-\infty}^{+\infty}dx'\,\mathcal{K}(x,t|x',t')\psi(x',t').
\end{equation}
From\refeq{propagator} and\refeq{evolution} we easily obtain~\cite{cufaro09}
\begin{equation*}
    i\partial_t\psi=-\frac{1}{\tau}\eta(\partial_x)\psi
\end{equation*}
where $\eta=\log\varphi$ and $\eta(\partial_x)$ is a pseudodifferential operator with symbol
$\eta(u)$~\cite{applebaum,cont,taylor,jacob} which plays the role of the generator of the
semigroup $T_t=e^{t\eta(\partial_x)/\tau}$ operating on a Banach space of measurable, bounded
functions~\cite{applebaum,cont,taylor,jacob}.

It is very well known~\cite{sato,applebaum}, on the other hand, that $\varphi$ represents an
infinitely divisible law if and only if $\eta(u)=\log\varphi(u)$ satisfies the L\évy--Khintchin
formula
\begin{equation}\label{LK}
    \eta(u)=i\gamma
    u-\frac{\beta^2u^2}{2}+\int_{\mathbb{R}}\left[e^{iux}-1-iux\,I_{[-1,1]}(x)\right]\,\nu(dx)
\end{equation}
where $\gamma,\beta\in\mathbb{R}$, $I_{[-1,1]}(x)$ is the indicator of $[-1,1]$, and $\nu(dx)$ is
a L\évy measure, namely a measure on $\mathbb{R}$ such that $\nu(\{0\})=0$ and
\begin{equation*}
    \int_{\mathbb{R}}(x^2\wedge1)\,\nu(dx)<+\infty.
\end{equation*}
In the most common cases of centered and symmetric laws the equation\refeq{LK} simplifies in
\begin{equation}\label{LKsymm}
    \eta(u)=-\frac{\beta^2u^2}{2}+\int_{\mathbb{R}}(\cos ux-1)\,\nu(dx)
\end{equation}
and $\eta(u)$ becomes even and real. As a consequence the corresponding operator
$\eta(\partial_x)$ is self--adjoint and acts on propagators and wave functions according to the
L\évy--Schr\"odinger integro--differential equation
\begin{equation}\label{lseq}
    i\partial_t\psi(x,t)=-\frac{1}{\tau}\,\eta(\partial_x)\psi(x,t)
                             =-\frac{\beta^2}{2\tau}\,\partial^2_x\psi(x,t)
                        -\frac{1}{\tau}\int_{\mathbb{R}}\left[\psi(x+y,t)-\psi(x,t)\right]\,\nu(dy)
\end{equation}
The integral term accounts for the jumps in the trajectories of the underlying stochastic process,
while an action $\alpha$ with $\beta^2=\alpha\tau/m$ weights the usual differential term of the
Schr\"odinger equation. For $\beta=0$ a pure jump L\évy--Schr\"odinger equation is obtained
\begin{equation}\label{symmlseq}
     i\partial_t\psi(x,t)=-\frac{1}{\tau}\int_{\mathbb{R}}\left[\psi(x+y,t)-\psi(x,t)\right]\,\nu(dy).
\end{equation}

\section{Stationary solutions for the free particle}

The free equation\refeq{lseq} admits simple stationary solutions: if we consider
\begin{equation*}
    \psi(x,t)=e^{-iE_0t/\alpha}\phi(x),\qquad\quad\alpha=\frac{m\beta^2}{\tau}
\end{equation*}
we then have
\begin{equation}\label{statsol}
    E_0\phi(x)=-\frac{\alpha^2}{2m}\,\phi''(x)-\frac{\alpha}{\tau}\int_\mathbb{R}[\phi(x+y)-\phi(x)]\,\nu(dy),
\end{equation}
and for a plane wave $\phi(x)=e^{\pm iux}$ from\refeq{LKsymm} -- namely with a symmetric L\évy
noise -- we get
\begin{equation*}
  E_0\phi(x) =-\frac{\alpha}{\tau}\left[-\frac{\beta^2u^2}{2}+\int_\mathbb{R}\left(e^{\pm iuy}-1\right)\,\nu(dy)\right]e^{\pm iux}
           =-\frac{\alpha}{\tau}\left[-\frac{\beta^2u^2}{2}+\int_\mathbb{R}(\cos uy-1)\,\nu(dy)\right]\phi(x)
           =-\frac{\alpha}{\tau}\eta(u)\phi(x)
\end{equation*}
which is satisfied when $E_0=-\alpha\eta(u)/\tau$. Hence, by taking $p=\alpha u$ as a momentum
variable, we obtain~\cite{cufaro09} the relevant equation
\begin{equation}\label{Ep}
    E_0=-\frac{\alpha}{\tau}\,\eta\left(\frac{p}{\alpha}\right)
\end{equation}
which connects the kinetic energy of a forceless particle to the logarithmic characteristic of a
L\évy process.

\section{Relativistic quantum mechanics}\label{rqm}

Let us take now in particular the non stable law
\begin{equation}\label{rqmeta}
    \eta(u)=1-\sqrt{1+a^2u^2}\,.
\end{equation}
With the following identification of the parameters
\begin{equation*}
    \alpha=\hbar,\qquad\quad\frac{\hbar}{\tau}=mc^2,\qquad\quad a=\frac{\hbar}{mc},\qquad\quad p=\hbar u.
\end{equation*}
we are led to the formula
\begin{equation}\label{relEp}
    E_0=-mc^2\eta\left(\frac{p}{\hbar}\right)=E-mc^2=\sqrt{m^2c^4+p^2c^2}-mc^2
\end{equation}
which is the well--known relativistic kinetic energy for a particle of mass $m$. The Schr\"odinger
equation of a relativistic free-particle is then easily obtained from\refeq{relEp} by
reinterpreting as usual $E$ and $p$ respectively as the operators $i\hbar\partial_t$ and
$-i\hbar\partial_x$
\begin{equation}\label{releq}
    i\hbar\partial_t\psi(x,t)=\sqrt{m^2c^4-\hbar^2c^2\partial^2_x}\,\psi(x,t),
\end{equation}
It is easy to check that this derives also from\refeq{lseq} after absorbing the mass energy term
$-mc^2$ of\refeq{relEp} into a phase factor $e^{imc^2t/\hbar}$. Remark that in three
dimensions\refeq{releq} would read
\begin{equation}\label{releq3}
    i\hbar\partial_t\psi(x,t)=\sqrt{m^2c^4-\hbar^2c^2\bm\nabla^2}\,\psi(x,t)
\end{equation}
It has been shown~\cite{applebaum,ichinose} that the L\évy process behind the
equations\refeq{releq} and\refeq{releq3} is a pure jump process~\cite{applebaum,cufaro09} with an
absolutely continuous L\évy measure $\nu(dx)=W(x)dx$ such that
\begin{equation}
    W(x)=\frac{1}{\pi|x|}\,K_1\left(\frac{|x|}{a}\right)=\frac{1}{\pi|x|}\,K_1\left(\frac{mc}{\hbar}\,|x|\right)
\end{equation}
($K_\nu$ are the modified Bessel functions~\cite{abramowitz}), that
in three dimensions becomes
\begin{equation}
    W(\bm x)=\frac{1}{2a\pi^2|\bm x|^2}\,K_2\left(\frac{|\bm x|}{a}\right)=\frac{mc}{2\hbar\pi^2|\bm x|^2}\,K_2\left(\frac{mc}{\hbar}\,|\bm x|\right)
\end{equation}
while from\refeq{symmlseq} the equation\refeq{releq} takes the form
\begin{equation}\label{intdiff}
    i\hbar\partial_t\psi(x,t)
    =-mc^2\int_{\mathbb{R}}\frac{\psi(x+y,t)-\psi(x,t)}{\pi|y|}\,K_1\left(\frac{mc}{\hbar}\,|y|\right)dy
\end{equation}
and in three dimensions is
\begin{equation}\label{intdiff3}
    i\hbar\partial_t\psi(\bm x,t)
    =-mc^2\int_{\mathbb{R}^3}\frac{\psi(\bm x+\bm y,t)-\psi(\bm x,t)}{2\pi^2|\bm y|^2}\,\frac{mc}{\hbar}K_2\left(\frac{mc}{\hbar}\,|\bm y|\right)d^3\bm
    y
\end{equation}
From the equation\refeq{releq3} -- by means of well known standard procedures~\cite{bjorken} --
one also derives (always for the free particle) the Klein--Gordon and Dirac equations in three
dimensions both for scalar, and for spinor wave functions $\psi$, namely respectively
\begin{eqnarray}
  \left(\square-\frac{m^2c^2}{\hbar^2}\right)\psi &=& 0, \label{KG}\\
  \left(i\gamma_\mu\partial^\mu-\frac{mc}{\hbar}\right)\psi  &=& 0.\label{D}
\end{eqnarray}
The corresponding Klein--Gordon and Dirac propagators in their turn satisfy the inhomogeneous
equations (here with $\hbar=c=1$)
\begin{eqnarray}
  \left(\square_2-m^2\right)\mathcal{K}_{KG}(2|1) &=& \delta^4(2|1)\label{propKG} \\
  \left(i\gamma_\mu\partial_2^\mu-m\right)\mathcal{K}_D(2|1) &=& i\,\delta^4(2|1)\label{propD}
\end{eqnarray}
with $\delta^4(2|1)=\delta(t_2-t_1)\delta^3(\bm x_2-\bm x_1)$. Let
us finally remark that these relativistic quantum wave equations
have been recently of particular interest~\cite{delgado} also in the
field of quantum optical phenomena and of quantum information.

\section{Infinite divisibility--preserving modifications}\label{modifications}

From the relativistic kinetic energy $E_0$ of a point particle of rest mass $m$
\begin{equation}\label{Tdimless}
    \frac{E_0(\bm p)}{mc^2}=\sqrt{1+\frac{\bm
    p^2}{m^2c^2}}-1.
\end{equation}
with the identifications
\begin{equation*}
    \eta=-\,\frac{E_0}{mc^2},\qquad\bm u=\frac{\bm p}{amc}
\end{equation*}
we obtain
\begin{equation*}
    \eta(\bm u)=1-\sqrt{1+a^2\bm u^2}
\end{equation*}
which of course coincides with the equation\refeq{rqmeta} extended to three dimensions. We take
now a class of transformations of $\eta(\bm u)$ characterized by the fact that they preserve the
infinite divisibility, while producing changes in the forceless particle equations of motion with
respect to the usual ones\refeq{KG} and\refeq{D}. To this end we modify the energy-momentum
formula as follows
\begin{equation}\label{etotmod}
    E(\bm p)=mc^2\sqrt{1+\frac{\bm
    p^2}{m^2c^2}+f\left(\frac{p^2}{m^2c^2}\right)}
\end{equation}
where $f$ is a -- possibly small -- dimensionless, smooth function of the relativistic scalar
$p^2/m^2c^2$. Of course this modification entails that $p^2$ no longer coincides with $m^2c^2$
since the standard energy-momentum relation is now changed into
\begin{equation}\label{pquadromod}
    p^2=\frac{E^2}{c^2}-\bm p^2=m^2c^2+m^2c^2f\left(\frac{p^2}{m^2c^2}\right).
\end{equation}
As we will see in the following, this also implies that the mass no longer is $m$: it will take
instead one or more values depending on the choice of $f$. As a matter of fact, it could appear to
be preposterous to introduce a function $f$ of an argument which after all is a constant (albeit
different from 1). However we will show that this artifice will lend us the possibility of having
both a mass spectrum, and a new wave equation when -- in the next section -- we will quantize our
classical relations. Moreover it will be argued in the following that we will choose $f$ in such a
way that the corresponding modified logarithmic characteristic $\eta$ will remain infinitely
divisible: a feature that is instrumental to keep a viable connection to a suitable underlying
L\évy process.

To see that, we first remark that\refeq{pquadromod} defines the total particle energy $E$ in an
implicit form. To find it explicitly we first rewrite\refeq{pquadromod} in a dimensionless form as
\begin{equation*}
    \frac{p^2}{m^2c^2}=1+f\left(\frac{p^2}{m^2c^2}\right),
\end{equation*}
and then, by taking $g(x)=x-f(x)$, we just observe that the former equation requires that $x$ be
solution of $g(x)=1$, namely
\begin{equation*}
    g\left(\frac{p^2}{m^2c^2}\right)=\frac{p^2}{m^2c^2}-f\left(\frac{p^2}{m^2c^2}\right)=1.
\end{equation*}
Note that $f$ and $g$ should be considered universal functions, and that the following conditions
hold
\begin{equation*}
    f(1)=0\qquad\qquad g(1)=1
\end{equation*}
If then $g^{-1}(1)$ represents one of the (possibly many) solutions
of this equation, we could write
\begin{equation*}
    \frac{p^2}{m^2c^2}=g^{-1}(1)=x(g)\Big|_{g=1}
\end{equation*}
so that we have
\begin{equation*}
    p^2=\frac{E^2}{c^2}-\bm p^2=m^2c^2g^{-1}(1)
\end{equation*}
which can be interpreted as a simple mass re-scaling, from $m$ to one of the (possibly many)
values $M=m\sqrt{g^{-1}(1)}$. The new hamiltonian then is
\begin{equation}\label{newhamilt}
    E(\bm p)=\sqrt{m^2c^4g^{-1}(1)+\bm p^2c^2}=Mc^2\sqrt{1+\frac{\bm
    p^2}{M^2c^2}}
\end{equation}
and its kinetic part (by applying the same re-scaling also to the subtracted rest mass term) is
\begin{equation*}
    E_0(\bm p)=E(\bm p)-mc^2\sqrt{g^{-1}(1)}
    =Mc^2\sqrt{1+\frac{\bm p^2}{M^2c^2}}-Mc^2.
\end{equation*}
Hence the main consequence of our modification consists of a re-scaling of the mass value ($m\to
M$) at a purely classical level. This fact is apparently welcome, because its straightforward
consequence is that the new associated logarithmic characteristic $\eta$ is trivially again
infinitely divisible, and hence still produces acceptable L\évy processes. But there is more:
since $g^{-1}(1)$ can take several different real and positive values, by means of our
modification\refeq{etotmod} we have introduced a mass spectrum: in the rest frame of the particle
we indeed have now
\begin{equation}\label{mass}
M=E_{cm}/c^2=m\sqrt{g^{-1}(1)}
\end{equation}

\section{Quantum equations of motion}

From the modified energy formula one derives a relativistic Schr\"odinger equation (for instance
by means of the formal substitutions $E\to i\hbar\partial_t$ and $\bm p\to-i\hbar\bm\nabla$):
\begin{equation}\label{releq3mod}
    i\hbar\partial_t\psi(x,t)= mc^2\sqrt{1-\frac{\hbar^2}{m^2c^2}\bm\nabla^2+f\left(\frac{\Box}{m^2c^2}\right)}\,\psi(x,t)
\end{equation}
where the square root pseudo-differential operator satisfies the constraints exposed in the
Section~\ref{intro}. From\refeq{releq3mod} one easily obtains in the usual manner (from here on
$\hbar=c=1$)
\begin{eqnarray}
  \left[\square-m^2f\left(\frac{1}{m^2}\,\square\right)-m^2\right]\psi &=& 0,\\
  \left[\square_2-m^2f\left(\frac{1}{m^2}\,\square_2\right)-m^2\right]\mathcal{K}_{KG}(2|1)\label{newpropKG}
  &=& \delta^4(2|1)
\end{eqnarray}
and, by standard methods~\cite{bjorken}, the \emph{modified} Dirac spinor equations
\begin{eqnarray}
  \left[i\gamma_\mu\partial^\mu-m\sqrt{1+f\left(\frac{1}{m^2}\,\square\right)}\,\right]\psi  &=& 0\\
  \left[i\gamma_\mu\partial_2^\mu-m\sqrt{1+f\left(\frac{1}{m^2}\,\square_2\right)}\,\right]\mathcal{K}_D(2|1)\label{newpropD}
  &=& i\delta^4(2|1)
\end{eqnarray}
In the momentum space (with Fourier transforms in four dimensions) these equations become much
simpler: more precisely we have
\begin{eqnarray}
  \mathcal{\mathcal{K}}_{KG}(p^2) &=& \frac{1}{p^2-m^2\left[1+f(p^2/m^2)\right]+i\epsilon}\label{prop1} \\
  \mathcal{\mathcal{K}}_D(p^2) &=& \frac{1}{\gamma^\mu
  p_\mu-m\sqrt{1+f(p^2/m^2)}+i\epsilon}\label{prop2}
\end{eqnarray}
We notice that $\mathcal{K}_D(2|1)$ is in our case simply related to the $\mathcal{K}_{KG}(2|1)$
(like in the usual case) as
\begin{equation}\label{prop3}
  \mathcal{K}_D(2|1)=i\left[\,i\dslash_2+m\sqrt{1+f(\square_2/m^2)}\,\,\right]\mathcal{K}_{KG}(2|1)
\end{equation}

\section{Phenomenology: quark and lepton masses}

The equations\refeq{newpropKG} and\refeq{newpropD} generalize the well known propagator
equations\refeq{propKG} and\refeq{propD} deriving from QED and QCD at zero order (in absence of
interaction terms). The Standard Model (SM)\footnote{For future developments we recall that the
Lagrangian density of QCD is, up to gauge fixing terms:
\begin{equation*}
    \mathcal{L}=-\frac{1}{4}F_{\mu\nu}^aF_a^{\mu\nu}+\sum_q\overline{\psi}_i^{\,q}[i\gamma^\mu(D_\mu)_{ij}-m_q\delta_{ij}]\psi_j^q
\end{equation*}
where $F_{\mu\nu}^a = \partial_\mu A^a_\nu-\partial_\nu
A^a_\mu+g_sf_{abc}A^b_\mu A^c_\nu$, and the insertion of interaction
terms is done with the minimal interaction by substituting the
simple derivative $\partial_\mu$ with the covariant one $D_\mu$
where we have respectively for QED and QCD
\begin{eqnarray*}
    D_\mu&\equiv&\partial_\mu-ieA_\mu \\
    \left(D_\mu\right)_{ij}&\equiv&\delta_{ij}\partial_\mu-ig_sT^a_{ij}A^a_\mu.
\end{eqnarray*}
Here $g_s$ is the QCD coupling constant, $T^a_{ij}$ and $f_{abc}$ are the $SU(3)$ color matrices
and structure constants respectively, and $A_\mu^a$ the eight Yang--Mills gluon fields; $\psi_i^q$
are the Dirac 4-spinors associated with each quark field of color $i$ and flavor $q$.}
$SU_c(3)\times SU_L(2)\times U(1)$ treats both strong, and electro-weak interactions: within this
scheme the modified $\eta(\bm u)$ leads to new interesting consequences. We begin by considering
the Feynman rules in perturbation theory in presence of the modified zero order propagator for
both spin $\frac{1}{2}$ (quarks and leptons) and spin $1$ (gluons, vector weak interacting
Bosons). The amplitude $A$ for a fermion that propagates from vertex $X$ to vertex $Y$, if
expanded, looks as follows: $A=A^{(0)}+A^{(1)}+A^{(2)}+\ldots$ The lowest order is
\begin{equation}
    A^{(0)}=Y\frac{i}{\gamma^\mu p_\mu-m\sqrt{1+f(p^2/m^2)}+i\epsilon}X.
\end{equation}
It is then possible that the Fermion emits and reabsorbs a virtual vector boson\footnote{The
presence of \emph{gammas} in the numerator of formulae\refeq{perturb} and\refeq{integral} is
typical of QED. More elaborated numerators may be present in non-abelian theories (in particular
QCD); however they appear totally unessential for our subsequent developments and purposes.} from
$X$ to $Y$:
\begin{eqnarray}\label{perturb}
    A^{(1)}&=&4\pi g_s^2Y\int d^4k\,\frac{\gamma^\mu}{\gamma^\rho
    p_\rho-m\sqrt{1+f(p^2/m^2)}}\,\frac{1}{(p-k)^2}\nonumber\\
          &&\qquad\qquad\qquad\times\frac{1}{k^\nu\gamma_\nu-m\sqrt{1+f(k^2/m^2)}+i\epsilon}\,\frac{\gamma_\mu}{\gamma^\rho p_\rho-m\sqrt{1+f(p^2/m^2)}}\,X
\end{eqnarray}
We choose now $f(x)$ in such a way that it makes finite the integral
\begin{equation}\label{integral}
    C=\gamma^\mu\int\frac{d^{\,4}k}{\gamma^\rho k_\rho-m\sqrt{1+f\left(k^2/m^2\right)}+i\epsilon}\,\frac{1}{(p-k)^2}\,\gamma_\mu
\end{equation}
One may notice that $f(x)$ behaves as a smooth \emph{cut-off} in a procedure of regularization at
each order in QCD (and QED). The integral $C$ is an invariant of the form
$C=\tilde{A}(p^2)\pslash+\tilde{B}(p^2)$ and its integrand is also present as a factor in higher
order terms, thus producing convergence.

The search of poles of the fermion propagators can be done in the following way~\cite{feynman}:
one considers the contributions of the perturbative expansion of the amplitude $A(p^2)$ (here we
always understand $f=f(p^2/m^2)$):
\begin{eqnarray}
  A(p^2) &=& Y\left\{\frac{1}{\pslash-m\sqrt{1+f}}+
  \frac{1}{\pslash-m\sqrt{1+f}}C\frac{1}{\pslash-m\sqrt{1+f}}\right.\nonumber\\
               && \qquad\qquad\qquad\qquad \left.+ \frac{1}{\pslash-m\sqrt{1+f}}C\frac{1}{\pslash-m\sqrt{1+f}} C\frac{1}{\pslash-m\sqrt{1+f}}
               +\ldots\right\}X
\end{eqnarray}
and using the formula
\begin{equation}
    \frac{1}{A-B}=\frac{1}{A}+\frac{1}{A}B\frac{1}{A}+\frac{1}{A}B\frac{1}{A}B\frac{1}{A}+\ldots
\end{equation}
one obtains the approximate expression
\begin{equation}\label{expansion}
    A\simeq Y\frac{1}{\pslash-m\sqrt{1+f}-C}X=Y\frac{1}{\pslash-m\sqrt{1+f}-\tilde{A}\pslash-\tilde{B}}X
\end{equation}
and looks for possible poles which -- after rationalizing equation\refeq{expansion} -- are
solutions of the equation
\begin{equation}
    \left[1-\tilde{A}(p^2)\,\right]^2p^2-\left[m\sqrt{1+f(p^2/m^2)}+\tilde{B}(p^2)\right]^2=0
\end{equation}

\subsection{Hypothesis for an approximate evaluation of the mass spectrum}

Let us now focus our attention on QCD. We may consider the approximation $\tilde{A}(p^2)\simeq
\tilde{A}(m^2)$ and $\tilde{B}(p^2)\simeq \tilde{B}(m^2)$ which follows from the assumption
\begin{equation}\label{}
     \tilde{A}\ll 1\qquad\mbox{and}\qquad\tilde{B}\ll m\sqrt{1+f(p^2/m^2)}
\end{equation}
We then obtain the equation
\begin{equation}
    p^2=\left[\frac{m\sqrt{1+f(p^2/m^2)}+\tilde{B}(m^2)}{1-\tilde{A}(m^2)}\right]^2=m^2_{exp}
\end{equation}
where the experimental masses $m_{exp}$ are still represented in an implicit form. However note
that in the limit $g_s\to0$, $\tilde{A}$ and $\tilde{B}\to0$ one achieves the equation
\begin{equation}\label{pquadro}
    p^2=m^2\left[1+f(p^2/m^2)\right]
\end{equation}
which coincides with the \emph{classical} equation\refeq{pquadromod}. At this point we notice that
the simplest choice of $f(x)$ that makes the integral $C$ finite (integrand convergent) is a
polynomial of third degree:
\begin{equation}\label{polynomial}
    f(x)=\lambda_0+\lambda_1x+\lambda_2x^2+\lambda_3x^3
\end{equation}
The equation\refeq{pquadro} then becomes
\begin{equation}\label{pquadro2}
    x-1-f(x)=-\lambda_3(x-1)(x-x_+)(x-x_-)=0\qquad\qquad \left(x=\frac{p^2}{m^2}\right)
\end{equation}
with
\begin{equation}\label{}
    f(1)=0\qquad\qquad\lambda_0=-\lambda_1-\lambda_2-\lambda_3
\end{equation}
where
\begin{equation}\label{}
    x_\pm=\frac{1}{2\lambda_3}\left(-\lambda_2-\lambda_3\pm\sqrt{\Delta}\right)\qquad\qquad\Delta=(\lambda_2-\lambda_3)^2-4\lambda_1\lambda_3-4\lambda_3^2+4\lambda_3
\end{equation}
and we finally get
\begin{eqnarray}
  \frac{\lambda_2}{\lambda_3} &=& -(1+x_++x_-) \\
  \frac{\lambda_1-1}{\lambda_3} &=& x_++x_-+x_+x_-
\end{eqnarray}
From the previous formulae one achieves the following interesting result: the convergence of $C$
determines a possible phenomenological function $f(x)$ that produces a mass spectrum of three
fermion particles (quarks in QCD).

The connections with the possible experimental physical masses are $M_1=m,\, M_2=m\sqrt{x_+},\,
M_3=m\sqrt{x_-}$. If the three poles in the free (zero order) propagator are real and positive
(with proper residues), with appropriate values of the $\lambda$'s, they allow the interpretation
of physical basic masses of fermions (quark or leptons) belonging to the three different families
of the Standard Model. To be more specific we get two different propagators for quarks, one with
charge $-\frac{1}{3}$ ($d,s,b$ quarks) and another with charge $+\frac{2}{3}$ ($u,c,t$ quarks).
Similarly for charged leptons (charge $-1$ and spin $\frac{1}{2}$) we get one propagator.

\section{Conclusions}
We have proposed a modification of the classical relativistic hamiltonian that allows the presence
of several masses without changing its basic structure. This modification does not affect the
infinite divisibility of the laws that are at the basis of the correspondence between stochastic
processes and L\'evy-Schr\"odinger equations. However we discovered that the mentioned
modification suggests a reformulation of the relativistic equations for wave functions and
propagators in such a way that a suitable choice of the background noise  produces a convergence
in the perturbative contributions. To this purpose  we remarked that a modification -- with
respect to the one given by equation\refeq{relEp} -- of the logarithmic characteristic $\eta(\bm
u)$, by the insertion of a cut-off $f(x)$, allows to proceed to regularization first, and then to
renormalization of the two-point function of QCD. There are three parameters in our
phenomenological $f(x)$ which is a third degree polynomial; the latter appears as the simplest
choice that produces convergence in the integrals representing high order contributions to the
fermion and boson propagators. Such parameters create three different poles in the zero-order
propagators and allow the interpretation of a physical system with three different masses under
precise constraints on $f(x)$. The masses might be related to the three families of the Standard
Model.

%\vfill \eject

\end{document}